\documentclass{vgtc}                          

\ifpdf
  \pdfoutput=1\relax                   
  \usepackage{graphicx}                
  \DeclareGraphicsExtensions{.pdf,.png,.jpg,.jpeg} 
\else
  \usepackage{graphicx}                
  \DeclareGraphicsExtensions{.eps}     
\fi%

\graphicspath{{figures/}{pictures/}{images/}{./}} 

\usepackage{microtype}                 
\PassOptionsToPackage{warn}{textcomp}  
\usepackage{textcomp}                  
\usepackage{mathptmx}                  
\usepackage{times}                     
\usepackage{cite}                      
\usepackage{tabu}                      
\usepackage{booktabs}                  

\ifpdf
  \pdfoutput=1\relax                   
  \usepackage{graphicx}                
  \DeclareGraphicsExtensions{.pdf,.png,.jpg,.jpeg} 
\else
  \usepackage{graphicx}                
  \DeclareGraphicsExtensions{.eps}     
\fi%

\graphicspath{{figures/}{pictures/}{images/}{./}} 

\usepackage{microtype}                 
\PassOptionsToPackage{warn}{textcomp}  
\usepackage{textcomp}                  
\usepackage{mathptmx}                  
\usepackage{mathtools}
\usepackage{times}                     
\usepackage{cite}                      
\usepackage{tabu}                      
\usepackage{booktabs}  
\usepackage{amsmath}
\usepackage[dvipsnames]{xcolor}

\onlineid{0}

\vgtccategory{Research}

\vgtcinsertpkg

\title{Opening Access to Visual Exploration of Audiovisual Digital Biomarkers: an OpenDBM Analytics Tool}

\author{Carla Floricel\thanks{e-mail: cflori3@uic.edu}\\ %
      \parbox{1.4in}{\scriptsize \centering University of Illinois Chicago} %
\and Jacob Epifano, Stephanie Caamano, Sarah Kark, Rich Christie, Aaron Masino, Andre D Paredes\thanks{e-mail:andre.paredes@aicure.com}\\ %
     \parbox{1.4in}{\scriptsize \centering AiCure}
}

\abstract{
Digital biomarkers (DBMs) are a growing field and increasingly tested in the therapeutic areas of psychiatric and neurodegenerative disorders. Meanwhile, isolated silos of knowledge of audiovisual DBMs use in industry, academia, and clinics hinder their widespread adoption in clinical research. How can we help these non-technical domain experts to explore audiovisual digital biomarkers? 
The use of open source software in biomedical research to extract patient behavior changes is growing and inspiring a shift toward accessibility to address this problem. OpenDBM integrates several popular audio and visual open source behavior extraction toolkits. We present a visual analysis tool as an extension of the growing open source software, OpenDBM, to promote the adoption of audiovisual DBMs in basic and applied research. Our tool illustrates patterns in behavioral data while supporting interactive visual analysis of any subset of derived or raw DBM variables extracted through OpenDBM.

} 

\begin{document}


\teaser{
  \centering
 \includegraphics[width=1\linewidth]{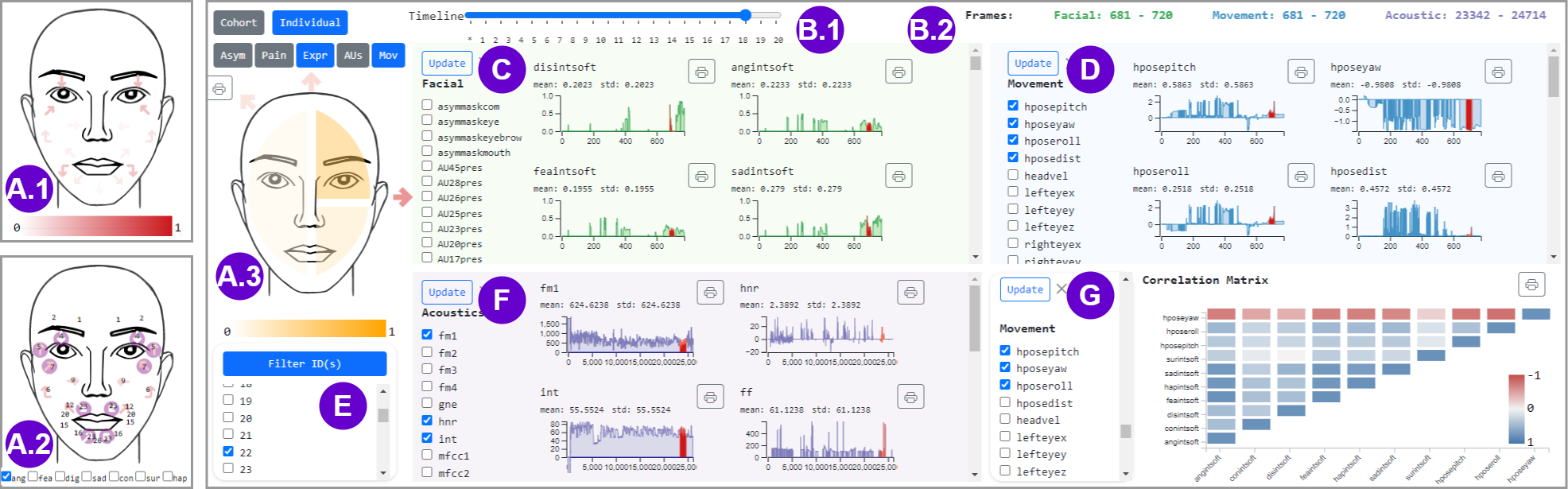}
\caption{Individual Panel. A) Head Sketch View which uses custom colored masks to display facial asymmetry, pain expressivity, overall expressivity (A.3), action unit (AU) intensity (A.1), and head movement (A.3) biomarkers. In A.2, the AUs' numbers are displayed and the AUs involved in anger expressivity (bottom selection) are marked with purple highlights. B.1) Timeline used to split the data into 20 time intervals. Upon change, the timeline will update the mean values displayed in A and highlight the selected interval in views C,D, and F, while the corresponding frame intervals will be shown in B.2. C, D, F) Facial, Movement, and Acoustics Views that display temporal data distributions for selected biomarker variables. E) ID Query Subpanel where one video can be chosen to display its DBM data. G) Correlation View that shows a pairwise Pearson correlation coefficients for a set of variables.}
\label{fig:individual}
} 
\maketitle

\section{Introduction} 
The global market value of digital biomarkers (DBMs) is forecasted to exceed \$7 billion by 2026\cite{bbcResearch}. DBMs are objective, quantifiable, physiological, and behavioral data collected and measured using digital devices. Like traditional biomarkers, DBMs have clinical value, such as predicting and diagnosing disease. However, DBMs introduce additional benefits that exceed traditional biomarkers' constraints, such as capturing longitudinal and continuous measurements that generate large, rich, and complex datasets\cite{babrak2019traditional}. Providing clinical researchers with practical tools to derive and interpret DBMs increases their ability to assess changes in health status relevant to healthcare applications\cite{vasudevan2022digital}. To support the rising demand for DBM adoption by clinical researchers, more practical tools are required to better inform non-technical biomedical researchers on how to use and identify DBMs\cite{h2020can,de2022digital}.

In particular, DBMs' increasing role in the therapeutic areas of psychiatric and neurodegenerative disorders provides newfound interest for clinical researchers to explore measurable audiovisual DBM changes to understand better how a patient feels \cite{benoit2020systematic, chandrabhatla2022co, jiang2022utilizing, kumar2022can, miller2022remote, piau2019current, pulido2020alzheimer, vsevvcik2022systematic, voleti2019review}. Growing open source software projects, such as OpenDBM, are lowering the barrier for non-technical clinical researchers to apply quantitative models, including machine learning models, to extract audiovisual features in human speech, voice acoustics, head movement, and facial expressions\cite{galatzer2020facial, galatzer2021validation}. However, despite accessible open source tools to extract audiovisual features, clinical researchers are burdened with interpreting large and complex quantitative datasets\cite{h2020can}. 

Given the novelty of DBMs and their still growing taxonomy and use\cite{coghlan2021digital}, there is interest among behavioral and biomedical researchers in finding practical tools that can facilitate exploratory analysis for data-informed hypothesis generation. This work aims to improve researchers' understanding of the breadth and scope of the hundreds of audiovisual DBMs available for investigatory adoption. We propose a visual analytics interface for the OpenDBM software\footnote{\texttt{https://aicure.com/opendbm/}}. Our proposed interface reveals patterns and outliers in facial, head movement, acoustics, and speech DBMs extracted from videos. To our knowledge, this work presents the first audiovisual DBM interactive visualization tool extracted from and made available through open source software.  


\section{Related Work and Background}

\noindent\textbf{Open Source Audio and Visual Feature Extraction Tools.} The system to measure the nature and intensity of vocal and facial expressions is advancing from manual raters to computerized toolkits \cite{rosenberg2020face, geiger2023computerized}. These audio and visual toolkits are made possible by leveraging advancements in machine learning and artificial intelligence techniques, such as natural language processing and computer vision\cite{corcoran2020using, jiang2022utilizing, bhadra2022insight, li2022applications}. 
A growing number of open source software projects are starting to make vocal and facial feature extraction toolkits freely available online. For vocal feature extractions, Parselmouth\cite{jadoul2018introducing}, Natural Language Toolkit\cite{loper2002nltk}, LexicalRichness\cite{lex}, and VaderSentiment\cite{yu2022automated} have been cited for calculating a whole host of speech and acoustic DBM variables. For facial feature extractions, OpenFace is a commonly cited behavior analysis toolkit for detecting and measuring facial landmarks, facial action units, head pose estimation, and gaze estimation\cite{baltruvsaitis2016openface, fydanaki2018evaluating, baltrusaitis2018openface}. Collectively, these software toolkits  provide a rich and diverse suite of extracted features for a more comprehensive analysis of emotional communication behavior over time. However, none of these projects provide visualization tools that can aid data interpretation. The project presented in this paper uses the OpenDBM solution, integrating all of the previously mentioned vocal and facial toolkits for generating a visualization of these extracted audiovisual features collectively.

\noindent\textbf{Visualization in Healthcare.} Visualization techniques in healthcare informatics often target cohort data exploration, covering applications in disease evolution from electronic medical records\cite{wongsuphasawat2011outflow, huang2015richly, wang2009temporal, wang2008aligning}, heterogeneous longitudinal clinical data\cite{harbig2021oncothreads, floricel2021thalis}, or volumetric patient data\cite{grossmann2019pelvis, wentzel2019cohort}. However, prior work in healthcare visual analytics mostly focus on chronic conditions such as cancer\cite{bernard2014visual}, stroke\cite{loorak2015timespan}, diabetes\cite{di2020s}, or on infectious disease control due to the COVID-19 pandemic\cite{baumgartl2020search, sondag2022visual}, and less on psychiatric and neurodegenerative disorders. Our work incorporates a new approach to promote individual patient data exploration, while incorporating past working approaches for cohort data exploration. There is work in visual analytics for facial activity and head movement and separately, for voice acoustics and speech measurements\cite{tifentale2015selfiecity, cheong2021py, mcduff2011acume, yamashita2019visualizing, ueng2012voice}, however, some of it doesn't use video data, and none accounts for all four measurement categories together. We aim to provide efficient tools for psychiatric and neurodegenerative health studies using heterogeneous, audiovisual, behavioral biomarker measurements extracted during clinical assessments.

\section{Design Process and Requirements}

The design process followed an Activity-Centered-Design approach\cite{marai2017activity}. Our team held remote meetings for nine weeks with five research groups in DBM therapeutic areas, collectively representing academia, clinics, and industry. While most collaborators were principal investigators with faculty positions, conducting behavioral or biomedical research, all of them were familiar with the OpenDBM software. Throughout this process, the team iteratively gained insight into user approaches to explore mappings between DBMs and conditions and disorders of interest (e.g., major depression and schizophrenia), gathered functional specifications for a DBM interface, and prototyped and evaluated the interface. Due to the large variety in patient behavior for these disorders, we gathered many specific requirements. However, we focused on the following subset of high-level requirements to serve all our collaborators and the open source community:

\textbf{R1:} Provide flexibility in showing details about any subset of DBM variables available through the OpenDBM pipeline. For instance, for early detection of Parkinson's disease, head movement measurements are of
greater importance than other DBM, such as voice acoustics. Adaptability to different workflows is an essential factor in open source. Additionally, analyzing hundreds of variables can be very challenging, and sometimes researchers don't know where to start their analyses. Thus, having the means and the freedom to choose what to explore visually is very important.


\textbf{R2:} Support interactive visualizations for both raw and derived data. Visualizing derived, mean variables is important for getting an effective overview of the cohort data and context for individual patients, while visualizing raw, temporal variables supports in-depth analysis for individual patients. This is critical for checking the quality of the data. For example, researchers might want to exclude from their analyses videos where the audio or the patient's face was not captured.

\textbf{R3:}  Emphasize trends and outliers in DBM data. As an example, patients should show negative emotions when talking about unpleasant or uncomfortable subjects. Domain experts should easily observe patterns between patients, which is helpful for further studies. Furthermore, highlighting correlations between biomarkers is fundamental for better understanding these conditions.

\section{Visualization Design}

The visual system is open source and can be operated through the OpenDBM Github project\footnotemark[2] from the visualization interface folder. It is not part of the DBM extraction pipeline, but serves as a complementary application that visualizes the output of the DBM extraction. The interface has two interactive panels: the Cohort Panel and the Individual Panel. These panels are composed of multiple coordinated views that support brushing and linking operations.

\subsection{Data Description}
Vocal and facial expressions convey emotion and communication behavior and are one of the most researched topics in psychology and related disciplines; as a result, audiovisual DBMs extend from these basic and applied science measurement tools \cite{geiger2023computerized}. When a video is processed through OpenDBM, the several vocal and facial feature extraction toolkits combine to present hundreds of unique variable categories relevant to four different audiovisual DBM  domains: speech, acoustics, facial expression, and head movement. Each audiovisual DBM domain provides two sets of quantitative variables: raw, captured as a frame-by-frame time sequence measurement, and derived, capturing summary statistics on the total collection of frames. These raw and derived variables provide a wide range of objective behavioral cues, such as transcription and lexical richness for speech, jitter and shimmer for acoustics, eye blink and facial tremor for head movement, and facial action units and facial asymmetry for facial expressions. The proposed interface uses these raw and derived variables to display relevant details and statistics about video cohorts and individual videos using two panels: the Cohort and the Invididual Panels. The official documentation\footnote{\label{note1}\texttt{https://github.com/AiCure/open\char`_dbm}} provides the full list of DBM variables extracted by OpenDBM.

\begin{figure*}[ht] 
\centering
\includegraphics[width=1\linewidth]{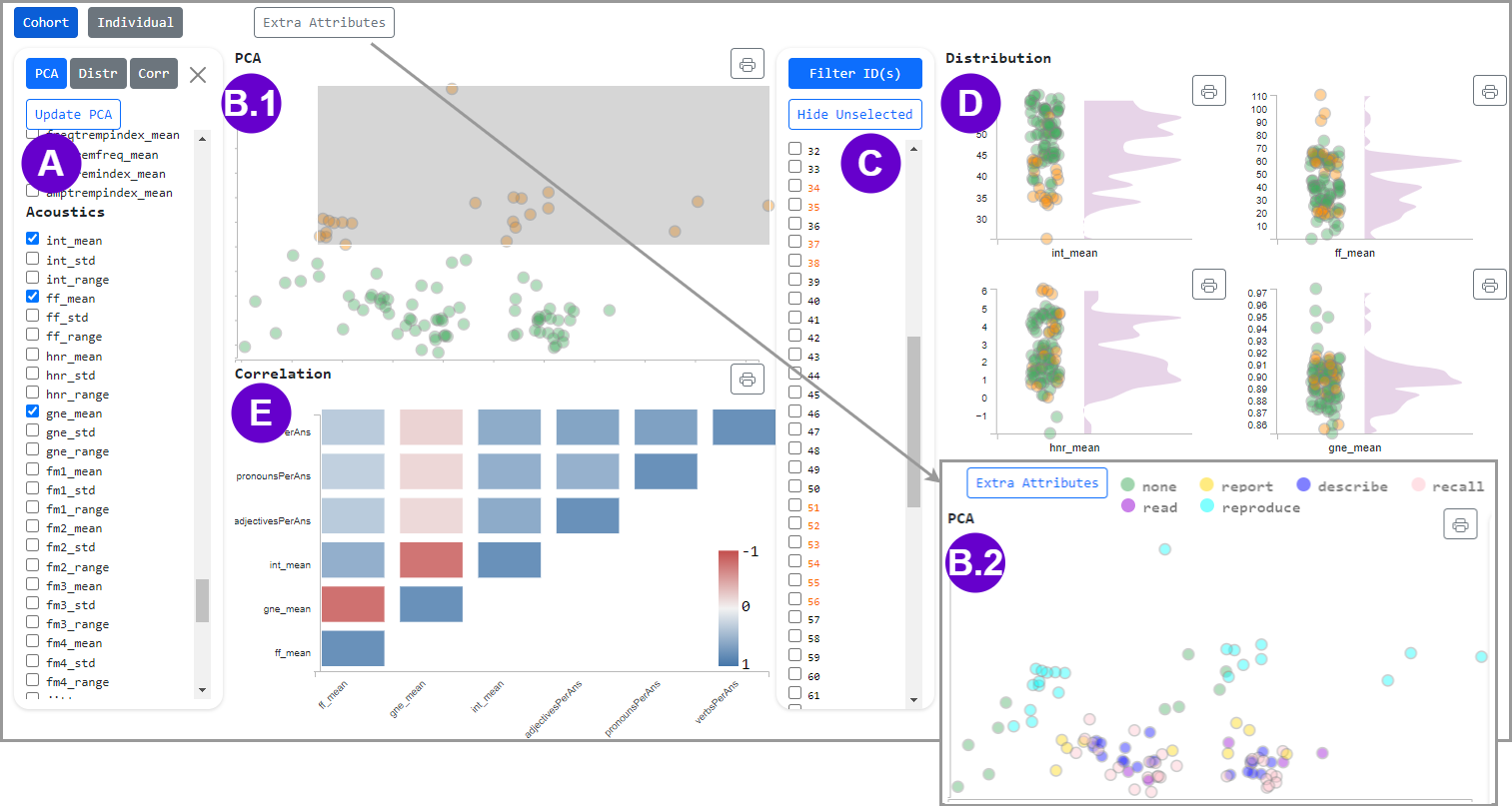}
\caption{Cohort Panel. A) Query Subpanel with three alternative components for biomarker variable selection for views B, D, E. B.1) PCA View that uses a scatterplot to display in 2D video data based on the variable selections in A.  B.2) The PCA scatterplot is color-coded based on an extra attribute, namely, the task that was performed during each video. C) ID Query Subpanel where IDs can be selected to be highlighted in views B and D, and unselected IDs can be hidden. D) Distribution View showing cohort distributions for four selected variables. E) Correlation View displaying pairwise Pearson correlation coefficients for six selected biomarker variables.}
\label{fig:cohort}
\end{figure*}

\subsection{Cohort Panel}
The Cohort Panel (Fig.~\ref{fig:cohort}) has three main views and functions: provide a cohort overview based on a selected set of variables, observe variable distributions, and find correlations between variables.

Two query subpanels are available for variable and video ID selection, with the variable query subpanel (Fig.~\ref{fig:cohort}.A) having three alternative components for each of the three main views (Fig.~\ref{fig:cohort}.B,D,E). In the video ID query subpanel (Fig.~\ref{fig:cohort}.C), selected IDs are highlighted in the other views while unselected videos can be hidden from the rest. All views have accompanying print buttons to generate plot images that can be used in further studies.

\noindent\textbf{PCA View.} This view (Fig.~\ref{fig:cohort}.B.1) uses a scatterplot for a cohort overview by arranging videos in 2D based on a selected set of biomarker variables (R1, R2, R3). The axes correspond to the first two components computed by a Principal Component Analysis (PCA) \cite{wold1987principal}. We employed factor analysis (i.e. PCA) to help researchers get a better sense of underlying structures in the high-dimensional video data while retaining patterns. The view shows trends and outliers, while brushing interactions highlight selected elements in the Distribution View and ID query subpanel. 

\noindent\textbf{Distribution View.} This view (Fig.~\ref{fig:cohort}.D) displays distribution charts for a selected set of biomarker variables (R1, R2, R3). Split into two components, each distribution chart shows one variable distribution across the cohort. The left side uses a scatterplot for easier detection of individual videos, while the right side uses a density plot for a concise cohort overview. Hovering on the scatterplot will highlight corresponding elements in the PCA Scatterplot and ID query subpanel and tooltips will display video IDs and variable values. 

Many times, domain experts study biomarker data while trying to find patterns between different cohorts. While gathering specifications for the system, a frequent request was to distinguish between subcohorts with different treatment plans, health conditions, age range, and so on. Thus, if such data is available, the system will color videos (Fig.~\ref{fig:cohort}.B.2) based on that list of extra attributes in both the PCA and Distribution Views (R3). 

\noindent\textbf{Correlation View.} This view (Fig.~\ref{fig:cohort}.E) contains a correlation matrix to emphasize interrelationships between a selected set of biomarker variables (R3). The matrix shows the coefficients computed using Pearson’s Correlation\cite{benesty2009pearson} method. Accompanying tooltips display the coefficient values for each pair or variables.

\subsection{Individual Panel}
This panel (Fig.~\ref{fig:individual}) has five coordinated views, showing temporal values for facial, movement, and acoustic variables, as well as a few derived variables, and correlations between raw variables. 

The panel features an ID query subpanel (Fig.~\ref{fig:individual}.E), where one video can be chosen to display its DBM data. An interactive timeline (Fig.~\ref{fig:individual}.B.1) splits the raw data into 20 time frames and highlights the corresponding time span in the other views upon change. Similar to the Cohort Panel, each view features print buttons.

\noindent\textbf{Head Sketch View.} This head sketch (Fig.~\ref{fig:individual}.A.1,2,3) supports derived value abstractions for facial activity and head pose biomarker data (R2, R3). Four alternative masks that use custom heatmaps can be applied to support facial biomarker visualization. The Asym mask uses colored glyphs to highlight facial features' asymmetry values. The Pain mask uses a colored glyph to highlight the pain expressivity values. The Expr mask uses colored glyphs to highlight the overall, upper, and lower face expressivity values (Fig.~\ref{fig:individual}.A.3). The AUs mask (Fig.~\ref{fig:individual}.A.1) highlights action units' intensity values using arrows pointing to the direction of the corresponding facial features' movements\cite{ekman1978facial}. As emotions are expressed through a combination of action units\cite{rosenberg2020face}, the AUs mask has a complementary layer (Fig.~\ref{fig:individual}.A.2) that indicates the set of AUs that get activated for each of the seven available emotions. Upon hovering, the action units’ numbers are visible on the facial sketch. Besides the facial activity masks, a head pose (Mov) mask (Fig.~\ref{fig:individual}.A.3) is also available and it indicates yaw, roll, and pitch head movements using colored pairs of arrows for each action. By default, the head sketch masks show derived values. However, this view will show mean values for the selected time frame when the timeline is updated.

\noindent\textbf{Facial, Movement, and Acoustics Views.} The Individual Panel features one view for temporal data for each of the three biomarker categories, namely, facial activity (Fig.~\ref{fig:individual}.C), head movement (Fig.~\ref{fig:individual}.D), and voice acoustics (Fig.~\ref{fig:individual}.F) (R1, R2, R3). Each view uses histograms to display temporal distributions for selected variables in the accompanying query subpanels, with the X axis representing time. The placement of these views facilitate the discovery of patterns among biomarkers. When the timeline is updated (Fig.~\ref{fig:individual}.B1), the corresponding time span is highlighted with red in all histograms while the frame spans are visible for each biomarker category (Fig.~\ref{fig:individual}.B.2).
 
\noindent\textbf{Correlation View.} This view (Fig.~\ref{fig:individual}.G) uses a correlation matrix to support the same functionality as the Correlation View in the Cohort Panel (R3). However, here, the Pearson Correlation is computed using longitudinal, individual data.

 DBMs, as an interdisciplinary tool in AI and medicine, have a far-reaching potential in basic and applied sciences that will continue to drive qualitative domain experts' interest in quantitative DBMs. Therefore, when we started this project, we realized the importance of making the interface accessible to a broad audience, including qualitatively trained domain experts. As a result, we used conventional visual encodings suitable for various visual literacy environments, such as scatterplots, histograms, density plots, and matrices. Additionally, we experimented with custom visualizations, such as the facial and head pose masks, taking advantage of the power of visual mapping techniques and trying to make better sense of the behavioral measurements from facial activity and head movement. We chose to first experiment with these two DBM categories because they were of interest in many use cases during our design requirements interviews. 

\section{Evaluation}We evaluated our system using two case studies that involved three domain experts and a video simulated actor dataset. Using simulated actor datasets is a well-accepted practice for capturing the prototypical representation of the multiple and complex emotional states of psychiatric and neurodegenerative disorders\cite{benda2020complex, cao2014crema, giuliani2017presentation, o2016eu}. A video stimulated actor dataset was generated and used for our case studies.  It included 95 videos, all under two minutes, of one adult male actor, instructed to perform five categories of tasks while simulating major depression disorder for some videos. The five tasks included: describe a picture, describe a memory, describe your day, read a passage, and reproduce a vowel sound or a basic facial emotional expression. The evaluation was conducted via video conferencing using screen sharing and the think aloud method.

\subsection{Case Study I: Cohort Analysis}
For this case study, the domain experts were first interested in discarding videos with no relevant information from the cohort (i.e., videos where the actor don't perform any tasks). The investigation started with the PCA View (Fig.~\ref{fig:cohort}.A, B.1), where audio intensity, fundamental frequency, and glottal to noise excitation ratio, all derived audio biomarker variables, were chosen as parameters to display the videos in 2D. Most videos were grouped towards the lower-central part of the view. Next, the previous set of parameters were used for the Distribution View (Fig.~\ref{fig:cohort}.D), and after brushing the upper part of the PCA scatterplot, the distribution charts revealed that the selected videos had the lowest audio intensity values from the cohort. When brushing the left or right outliers, the distribution charts showed that the selections belonged to opposite cohort extremities for fundamental frequency and glottal to noise excitation ratio, which aligned with previous research\cite{michaelis1997glottal}. When using the option to color the videos by the task category (Fig.~\ref{fig:cohort}.B.2), the scatterplot showed that the upper and left sides of the view was composed of videos with no audio or videos where the actor reproduced facial expressions or vowels. Lastly, the domain experts wanted to check the correlations between these parameters and speech DBM variables, such as the number of pronouns, verbs, or adjectives used per task (Fig.~\ref{fig:cohort}.E). The Correlation View revealed strong positive correlations between speech DBM variables and the audio intensity. This case study showed the interface's ability to show relevant trends and outliers in subsets of derived DBM variables of interest (R1, R2, R3) and helped the evaluators to detect videos with no acoustic DBM data.

\subsection{Case Study II: Individual Analysis}
This case study started with the exploration of raw data for a video where the actor performs a picture description task (Fig.~\ref{fig:individual}.E). The facial emotion expressions were chosen as parameters for the Facial View (Fig.~\ref{fig:individual}.C). The histograms showed high spikes for most negative emotions, such as distress, anger, fear, and sadness, implying that the task entailed describing a negative impact image. Hence, the domain experts were interested in observing facial action unit changes over time, since they are connected to facial expressions, so they used the Head Sketch View (Fig.~\ref{fig:individual}.A.1) and the timeline for this task (Fig.~\ref{fig:individual}.B.1). After applying the AUs facial mask for action units and the Mov mask for head pose changes (Fig.~\ref{fig:individual}.A.3), they investigated different time frames using the timeline, and observed that, indeed, at most times, the action units for negative emotions were active (Fig.~\ref{fig:individual}.A.2). Next, the evaluators checked correlations between head poses and emotions using the Correlation View (Fig.~\ref{fig:individual}.G), and observed that, surprisingly, roll head movements were negatively correlated with most emotions, while the other head poses didn’t show any particularly strong correlations to emotions. Curious, the evaluators watched the actual video, which showed the actor describing a picture of a building on fire. This case study showcases the interface's ability to show patterns in selected raw DBM variables for one video (R1, R2, R3) and helped the evaluators to detect a video where negative emotional impact was present.


\section{Discussion and Conclusion}



We received positive feedback for the current version of our interface in particular because, previously, domain experts were limited to manual and laborious means of inspecting video data: \textit{``Very cool, so much better to use for the analysis we did last year, huge time saver,"} When asked what they found most useful for their own research studies, most people pointed out to the interface's ability to delineate subcohorts using different colors (Fig.~\ref{fig:cohort}.B.2) \textit{``very excited about the color coding,"} as well as the histograms (Fig.~\ref{fig:individual}.C,D,F) \textit{``I had someone looking away from the camera, this is actually picking up their data."} However, one evaluator indicated that it would be useful to \textit{``input our own data [in the interface] to work with the [bio]markers,"} as it could speed up symptom research for different disorders. We plan to address this matter in our future work.

Overall this is a well-received start toward introducing visualization approaches to domain experts in DBM therapeutic fields, but it is still a work in progress. There are several limitations to our current solution, which we plan to improve. For example, the Cohort Panel suffers from scalability issues (i.e., the scatterplots) when it comes to large cohorts of hundreds of videos. Furthermore, we are currently researching missing DBM normative ranges, which better inform domain experts on abnormal patterns and help generate more accurate hypotheses about an individual's health status. 


This visualization approach begins to address the need for transparency behind data quality control and quality assurance from these integrated open source toolkits. The open-source space is a natural fit to drive domain expert users to test and ratify best practices. Future work can build upon our visualization approach by further improving visualizations of raw data for checking data quality from new and unvalidated toolkits. 

We conclude that our work offers the first visualization analysis tool for domain experts interested in exploring and using audiovisual DBMs, meeting the following five criteria: (1) hundreds of audio and visual DBM variables (2) cohort visualization panel to identify patterns in groups of audiovisual recordings (3) individual panel to identify single patient patterns in vocal and facial expression over time (4) open-source software  (5) integration to a growing number of open source audio and visual feature extraction toolkits, via OpenDBM.
\bibliographystyle{abbrv-doi}

\bibliography{template}
\end{document}